# Benchmarking GPT-5 for Zero-Shot Multimodal Medical Reasoning in Radiology and Radiation Oncology

Mingzhe Hu[†,1], Zach Eidex[†,1], Shansong Wang[1], Mojtaba Safari[1], Qiang Li[1], and Xiaofeng Yang[1*]

[1]Department of Radiation Oncology and Winship Cancer Institute, Emory University, Atlanta, GA

*Corresponding to: xiaofeng.yang@emory.edu

†These authors contributed equally to this work.

**Abstract**

Radiology, radiation oncology, and medical physics demand decision-making that integrates medical images, textual reports, and quantitative data under high-stakes conditions. With the introduction of GPT-5, it is critical to assess whether recent advances in large multimodal models translate into measurable gains in these safety-critical domains. We present a targeted zero-shot evaluation of GPT-5 and its smaller variants (GPT-5-mini, GPT-5-nano) against GPT-4o across three representative tasks: (1) VQA-RAD, a benchmark for visual question answering in radiology; (2) SLAKE, a semantically annotated, multilingual VQA dataset testing cross-modal grounding; and (3) a curated Medical Physics Board Examination-style dataset of 150 multiple-choice questions spanning treatment planning, dosimetry, imaging, and quality assurance. Across all datasets, GPT-5 achieved the highest accuracy, with substantial gains over GPT-4o—up to +20.00% in challenging anatomical regions such as the chest-mediastinal, +13.60% in lung-focused questions, and +11.44% in brain-tissue interpretation. On the board-style physics questions, GPT-5 attained 90.7% accuracy (136/150), exceeding the estimated human passing threshold, while GPT-4o trailed at 78.0%. These results demonstrate that GPT-5 delivers consistent and often pronounced performance improvements over GPT-4o in both image-grounded reasoning and domain-specific numerical problem-solving, highlighting its potential to augment expert workflows in medical imaging and therapeutic physics.

## 1 Introduction

Radiology, radiation oncology, radiation oncology and medical physics are domains in which decision-making is fundamentally tied to the integration of multimodal information. Clinical workflows in these fields routinely require joint interpretation of medical images, free-text reports, and quantitative parameters—such as Hounsfield units, dose-volume histograms, or geometric uncertainties—under constraints that are often urgent and high-stakes. In radiology, accurate image-based reasoning directly informs diagnosis, while radiation oncology must extend such reasoning into spatially precise treatment design. Medical physicists contribute specialized domain knowledge to ensure the safety, accuracy, and reliability of imaging and therapeutic systems. As such, AI systems intended to support these domains must be able to handle domain-specific logic, integrate structured and unstructured inputs, and reason over images in clinically coherent ways[1].

Earlier generations of general-purpose LLMs demonstrated that strong baseline performance could be achieved on specialized clinical tasks without large volumes of task-specific training data[2]. Models such as GPT-3.5 and GPT-4 proved capable of answering complex medical questions in zero-shot and few-shot settings, following detailed instructions, and adapting to varied contexts through conversational interaction[3]. These capabilities have been explored across a broad spectrum of specialties, from surgical domains such as neurosurgery to internal medicine subspecialties and disease-specific areas such as hepatology[4]. In parallel, early investigations in radiology, pathology, and orthodontics highlighted the potential for such models to assist with image interpretation, diagnostic reasoning, and treatment planning. Beyond research settings, LLMs have been tested in operational workflows for drafting clinic notes, summarizing discharge records, and generating preventive care recommendations[5-7].

Despite this progress, the release of GPT-5 introduces a significant opportunity to revisit the question of model capability—especially in multimodal and safety-critical domains such as radiology and radiation oncology—under a more advanced modeling framework. Although GPT-5 has shown remarkable improvements on general-domain[8] and professional reasoning benchmarks[9], its performance in radiology- and oncology-specific tasks remains untested. In particular, it is unclear whether such improvements extend to tasks that require tight image-text integration, or to board-style physics questions that demand precise numeric understanding and domain-specific reasoning. This motivated us to conduct a targeted evaluation of GPT-5's zero-shot performance across three curated tasks that reflect core challenges in medical imaging and treatment science.

Our study evaluates GPT-5 on (1) VQA-RAD[10], a benchmark for visual question answering in radiology, assessing how well the model can extract findings from clinical images and reason over visual-textual contexts; (2) SLAKE[11], a multilingual, semantically annotated VQA dataset designed to test cross-modal grounding and concept-level understanding; and (3) a carefully constructed set of Medical Physics Board Examination-style questions, representative of high-stakes testing scenarios in radiation oncology and therapeutic physics. All tasks are evaluated in a zero-shot setting using standardized prompts and formats, without domain-specific fine-tuning or retrieval. This protocol allows us to isolate the generalization capacity of GPT-5 under realistic, clinically inspired task conditions.

By combining image-grounded clinical reasoning with structured exam-style assessments, this work aims to provide the first comprehensive snapshot of GPT-5's potential—and limitations—in radiology and radiation oncology–oriented multimodal intelligence. In doing so, we hope to inform both near-term applications in clinical decision support and longer-term efforts to adapt foundation models for high-reliability, imaging-driven medical disciplines.

## 2 Methodology

### 2.1 Datasets

To evaluate the performance of GPT-5 on multimodal medical reasoning in radiology and radiation oncology, we selected three representative datasets: VQA-RAD[10], SLAKE[11], and a curated collection of Medical Physics Board Exam–style questions. Each dataset captures different aspects of clinical reasoning, ranging from image-grounded question answering to domain-specific knowledge assessment.

VQA-RAD is a visual question answering dataset derived from real-world radiology images and radiology report–style questions. It contains binary yes/no questions paired with chest, abdominal, and musculoskeletal images, requiring models to understand visual content in clinical context. We use the official test split under the Test-CSS protocol, which contains 251 question–image pairs. The dataset is well-suited to evaluate image-based factual reasoning in radiology without dependence on extended medical narratives.

SLAKE (Semantically Labeled Knowledge-Enriched medical VQA dataset) is a multilingual benchmark designed to evaluate multimodal comprehension and semantic alignment. It includes medical images and multiple-choice questions labeled with detailed knowledge components such as anatomical site, modality, and disease category. We follow the Test-CSS split, which ensures standardized evaluation and avoids data leakage. SLAKE introduces more diverse image types and question structures than VQA-RAD, testing the model's robustness across languages and semantic categories.

Medical Physics Board Exam is a dataset curated in-house, consisting of 150 high-quality multiple-choice questions modeled after standardized certification exams in radiation oncology physics. These questions cover a wide range of topics, including treatment planning systems, dosimetry, beam modeling, image-guided radiation therapy, quality assurance, and radiation protection[12]. All items contain four or five answer choices, with a single correct answer. The dataset is designed to test conceptual clarity, numerical estimation, and theoretical understanding relevant to clinical physics practice.

### 2.2 Prompting Design

Each sample was formulated as a single interaction, consisting of a system message and a user message. The system message was tailored to the specific domain of each dataset, while the user message encoded the question, image (if applicable), and answer format. The model was required to directly output the final answer without any intermediate explanation or reasoning chain.

We adopted a unified zero-shot prompting strategy across all evaluations, using a single-turn interaction without in-context exemplars, step-by-step reasoning, or retrieval assistance. For each dataset, the prompt consisted of a system message tailored to the domain and a user message containing the question and, when applicable, the corresponding image. The model was required to output one final answer directly, without generating intermediate rationale or explanation.

For VQA-RAD, which includes both binary yes/no and open-ended free-text questions paired with clinical radiology images, the prompt format was adapted to the nature of the question. The system message instructed the model to act as *"a diagnostic radiologist"* and respond accordingly: *"Use the provided medical image to answer each question as clearly and clinically as possible. Respond with 'Yes' or 'No' if the question is binary, or with a short medical phrase when open-ended."* This design allowed the model to produce succinct and context-appropriate answers based on image interpretation, reflecting real-world radiological decision-making.

For SLAKE, which contains both multiple-choice and open-ended questions grounded in annotated clinical images, the prompt was configured to support semantic understanding and visual reasoning. The system message defined the model as *"a medically trained AI assistant specialized in interpreting annotated clinical images"*, with further guidance: *"Carefully review each image and question, and respond with either the most accurate answer choice or, if no options are given, a concise and conceptually meaningful phrase that reflects the clinical interpretation."* For multiple-choice items, all options labeled A through D (or E) were included in the user message. For open-ended items, the model was expected to generate a short, medically meaningful response based on the image content.

For the Medical Physics Board Exam, which consists entirely of multiple-choice, text-only questions, the prompt followed a formal assessment style. The model was instructed to act as *"a board-certified medical physicist"*, with the system message: *"Please select the most appropriate answer based on the provided options."* The question and choices were presented in structured exam format, and the model was expected to choose a single correct answer from A through D or E.

This prompting design ensures consistency across datasets while preserving the unique characteristics of each task. All outputs were evaluated based solely on the final answer, with no intermediate reasoning steps permitted.

## 3 Results

### 3.1 Overall and Answer-Type-Specific Performance on SLAKE

We first evaluate the performance of four GPT models—GPT-5, GPT-5-mini, GPT-5-nano, and GPT-4o—on the SLAKE dataset across different answer types. As shown in Figure 1 and Table 1, GPT-5 achieves the highest accuracy across all categories, including aggregate, open-ended, and closed-ended responses. Specifically, GPT-5 attains an aggregate accuracy of 88.60%, representing a relative improvement of +14.78% over GPT-4o. This performance advantage is even more pronounced for open-ended questions, where GPT-5 outperforms GPT-4o by +16.31%, highlighting its superior generation capabilities in free-form medical image question answering. In contrast, GPT-5-nano and GPT-4o demonstrate notably lower accuracy, especially in open-ended settings, indicating their limited reasoning capacity under generative conditions. Even in closed-ended formats, where simpler models typically perform better, GPT-5 still maintains the lead with 92.31%, surpassing GPT-4o by +12.61%. These results underscore the consistent superiority of GPT-5 in both constrained and unconstrained response scenarios.

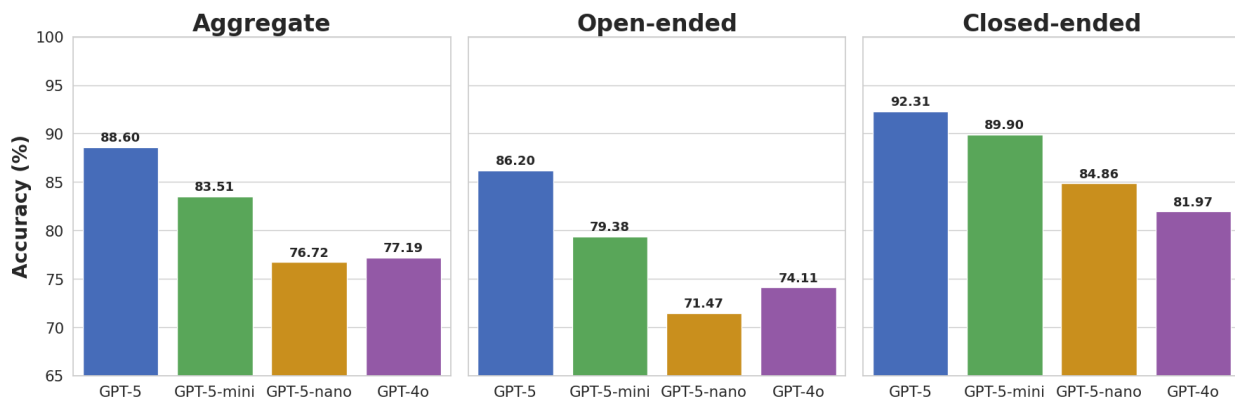

**Figure 1.** Comparative accuracy of four GPT models—GPT-5, GPT-5-mini, GPT-5-nano, and GPT-4o—on the SLAKE dataset. The results are presented across three answer types: (1) Aggregate Accuracy (left), representing overall performance across all question types; (2) Open-ended Accuracy (middle), focusing on free-form answer generation; and (3) Closed-ended Accuracy (right). GPT-5

consistently outperforms other variants across all categories, while smaller models like GPT-5-nano and GPT-4o show reduced accuracy, especially in open-ended settings.

**Table 1.** Model performance comparison across answer types on the SLAKE dataset. Accuracy values are reported for four GPT models—GPT-5, GPT-5-mini, GPT-5-nano, and GPT-4o—across three answer types: aggregate (all questions), open-ended, and closed-ended. The "↑X%" values in the GPT-5 column indicate the relative improvement of GPT-5 over GPT-4o, calculated as the percentage increase in accuracy compared to GPT-4o. GPT-5 consistently outperforms GPT-4o, with the most notable gain observed in open-ended questions (+16.31%).

| Answer Type | GPT-5 (vs 4o) | GPT-5-mini | GPT-5-nano | GPT-4o |
|---|---|---|---|---|
| Aggregate | **88.60%** (↑14.78%) | 83.51% | 76.72% | 77.19% |
| Open-ended | **86.20%** (↑16.31%) | 79.38% | 71.47% | 74.11% |
| Closed-ended | **92.31%** (↑12.61%) | 89.90% | 84.86% | 81.97% |

### 3.2 Content Type Understanding

Figure 2 and Table 2 present the classification accuracy across ten distinct content types commonly encountered in medical visual question answering tasks. These include structural categories such as Modality, Organ, Plane, and Position, as well as semantic categories like Abnormality, Quantity, Shape, and Size. GPT-5 achieves the highest accuracy in eight out of ten content types, with the most substantial performance lead observed in Color (+43.52%), Position (+45.35%), and Size (+38.12%) compared to GPT-4o. These gains suggest that GPT-5 exhibits stronger multimodal reasoning abilities, particularly in tasks requiring fine-grained visual and spatial comprehension. While GPT-5-mini and GPT-5-nano occasionally match GPT-5 in isolated categories (e.g., Shape and Plane), their overall performance remains lower, especially in conceptually demanding tasks such as KG-based and Position-related questions. This indicates that model scale plays a pivotal role in accurately interpreting complex visual semantics in medical images.

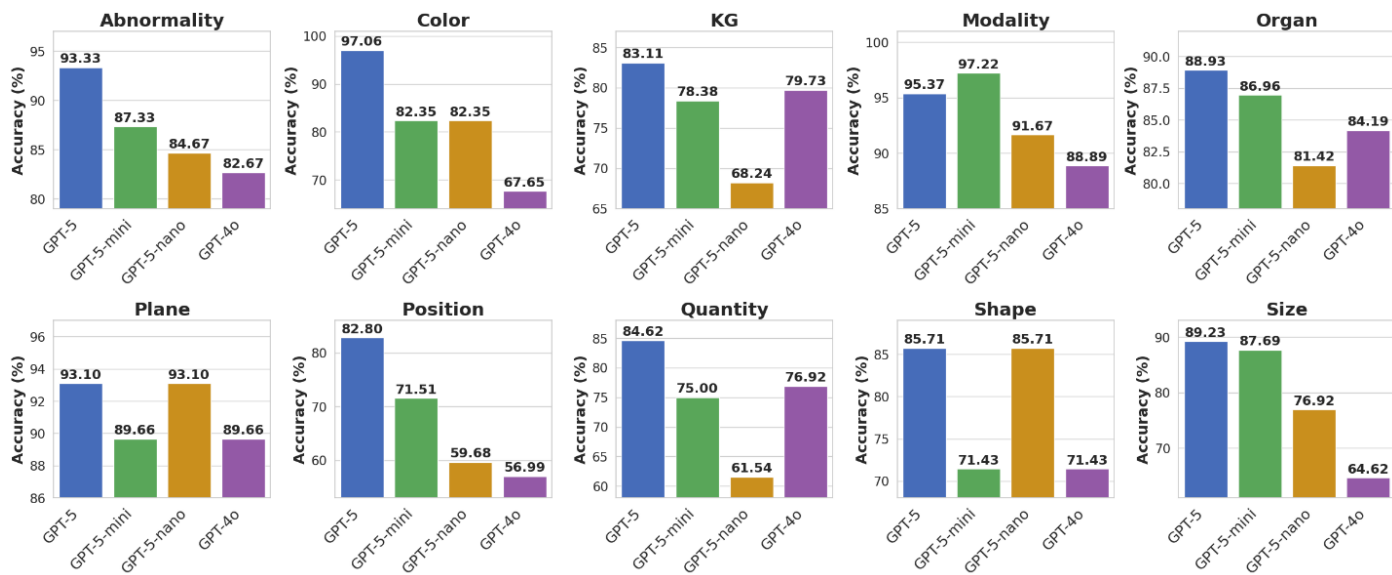

**Figure 2.** Accuracy of four GPT variants (GPT-5, GPT-5-mini, GPT-5-nano, and GPT-4o) across ten medical image content types. The categories include Abnormality (e.g., lesions, fractures), Color (e.g., hyperintensity), KG (knowledge-grounded questions requiring external clinical or commonsense information), Modality (e.g., CT, MRI), Organ (e.g., lung, brain), Plane (e.g., axial, sagittal), Position (e.g., left, anterior), Quantity (e.g., number of lesions), Shape (e.g., round, irregular), and Size (e.g., enlarged, small). Each subfigure reports the classification accuracy for the respective content type.

**Table 2.** Comparison of content type classification accuracy across four models—GPT-5, GPT-5-mini, GPT-5-nano, and GPT-4o—on multiple medical visual question answering categories. The "↑X%" values in the GPT-5 column indicate the relative improvement of GPT-5 over GPT-4o, calculated as the percentage increase in accuracy compared to GPT-4o.

| Content Type | GPT-5 (vs 4o) | GPT-5-mini | GPT-5-nano | GPT-4o |
|---|---|---|---|---|
| Abnormality | **93.33%** (↑12.94%) | 87.33% | 84.67% | 82.67% |
| Color | **97.06%** (↑43.52%) | 82.35% | 82.35% | 67.65% |
| KG | **83.11%** (↑4.23%) | 78.38% | 68.24% | 79.73% |
| Modality | 95.37% (↑7.29%) | **97.22%** | 91.67% | 88.89% |
| Organ | **88.93%** (↑5.63%) | 86.96% | 81.42% | 84.19% |
| Plane | **93.10%** (↑3.83%) | 89.66% | **93.10%** | 89.66% |
| Position | **82.80%** (↑45.35%) | 71.51% | 59.68% | 56.99% |
| Quantity | **84.62%** (↑10.00%) | 75.00% | 61.54% | 76.92% |
| Shape | **85.71%** (↑20.00%) | 71.43% | **85.71%** | 71.43% |
| Size | **89.23%** (↑38.12%) | 87.69% | 76.92% | 64.62% |

### 3.3 Region-Specific Accuracy Comparison

To further dissect the model's performance across different anatomical contexts, we evaluated classification accuracy over eight predefined regions of interest (ROIs), including the abdomen, brain, craniofacial region, brain tissue, chest-mediastinal, lung, neck, and pelvic cavity (Figure 3 and Table 3). The results reveal that GPT-5 consistently outperforms all other evaluated models across the majority of anatomical regions, reinforcing its robustness in domain-specific visual question answering (VQA) tasks.

GPT-5 achieved the highest accuracy in all eight ROIs, with particularly strong performance in the lung (94.03%), chest-mediastinal (90.00%), and brain (90.48%) regions. These areas are often characterized by complex anatomical structures and subtle diagnostic cues, underscoring GPT-5's advanced capacity for fine-grained reasoning. Notably, GPT-5 achieved a 20.00% relative improvement over GPT-4o in the chest-mediastinal region, followed by substantial gains in the lung (↑13.60%) and brain tissue (↑11.44%).

**Figure 3.** Comparison of model accuracy across different regions of interest (ROIs), including brain, lung, abdomen, pelvic cavity, neck, and chest-mediastinal. GPT-5 consistently outperforms other models across most ROIs, showing particularly strong performance in thoracic and head-neck regions. While GPT-5-mini achieves comparable accuracy in certain regions such as neck and chest-mediastinal, GPT-5-nano and GPT-4o exhibit greater variability, especially in complex anatomical areas.

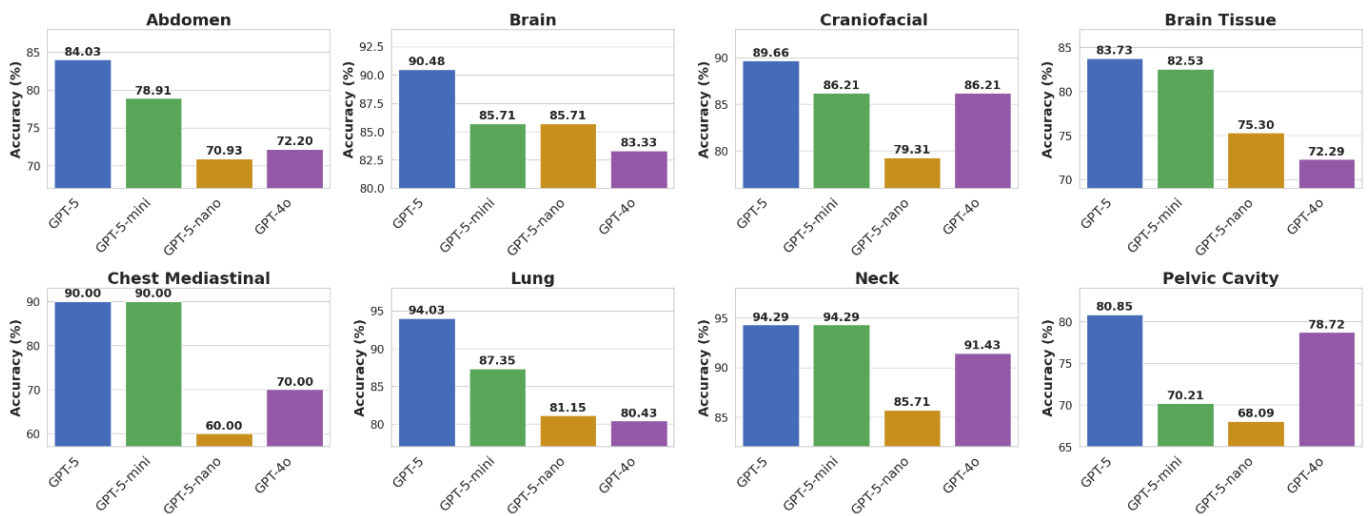

| ROI Region | GPT-5 (vs 4o) | GPT-5-mini | GPT-5-nano | GPT-4o |
|---|---|---|---|---|
| Abdomen | **84.03% (↑11.83%)** | 78.91% | 70.93% | 72.20% |
| Brain | **90.48% (↑7.15%)** | 85.71% | 85.71% | 83.33% |
| Brain_Face | **89.66% (↑3.45%)** | 86.21% | 79.31% | 86.21% |
| Brain_Tissue | **83.73% (↑11.44%)** | 82.53% | 75.30% | 72.29% |
| Chest-mediastinal | **90.00% (↑20.00%)** | **90.00%** | 60.00% | 70.00% |
| Lung | **94.03% (↑13.60%)** | 87.35% | 81.15% | 80.43% |
| Neck | **94.29% (↑2.86%)** | **94.29%** | 85.71% | 91.43% |
| Pelvic Cavity | **80.85% (↑2.13%)** | 70.21% | 68.09% | 78.72% |

**Table 3.** Accuracy comparison of GPT models across different anatomical regions of interest (ROIs). GPT-5 consistently outperforms other models across most regions, with especially pronounced improvements in the chest-mediastinal (↑20.00%) and lung (↑13.60%) regions. GPT-5-mini shows comparable performance in the neck and chest-mediastinal areas, while GPT-5-nano and GPT-4o exhibit more variability, particularly in abdominal and pelvic cavity segmentation. Accuracy values are reported in percentage, and performance gains of GPT-5 over GPT-4o are indicated in parentheses.

### 3.3 Modality-wise Performance Comparison

We further examined the performance of all four GPT-based models across different imaging modalities, including computed tomography (CT), magnetic resonance imaging (MRI), and X-ray (Figure 4 and Table 4). GPT-5 consistently outperformed its counterparts across all modalities, demonstrating the strongest robustness and generalization capability. In particular, GPT-5 achieved the highest accuracy in X-ray interpretation (94.74%), representing a substantial gain of 14.41% over GPT-4o. This indicates GPT-5's superior capacity for handling high-noise, low-contrast radiographs, which often present challenges for earlier models.

For MRI, GPT-5 reached an accuracy of 84.65%, outperforming GPT-4o by 10.97%. Although GPT-5-mini also achieved competitive performance (82.89%), both GPT-5-nano and GPT-4o fell short in this modality, likely due to the inherent complexity and variation in MRI contrast characteristics across sequences and anatomical structures.

In CT scans, GPT-5 maintained its leading position with an accuracy of 85.81%, showing a notable improvement of 9.33% over GPT-4o. The performance gap was more evident when comparing with GPT-5-nano (74.79%) and GPT-4o (76.48%), suggesting that smaller model variants struggle with the structural intricacies and subtle intensity variations in CT data.

These results collectively indicate that GPT-5 not only scales well with input complexity but also demonstrates cross-modality adaptability, maintaining state-of-the-art performance across diverse imaging domains.

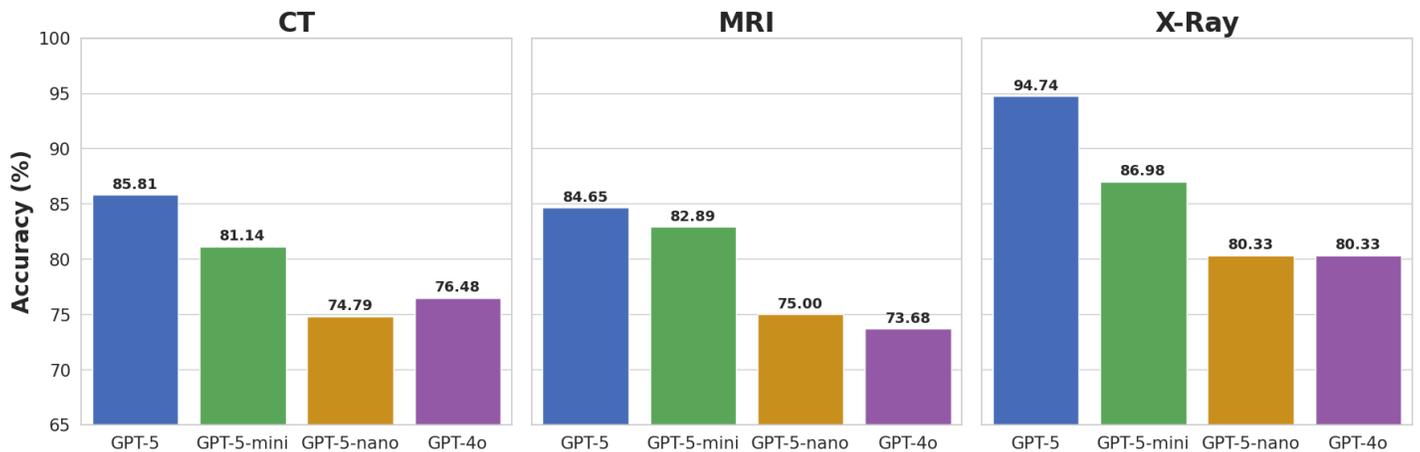

**Figure 4.** Comparison of model accuracy across different imaging modalities, including CT, MRI, and X-ray. GPT-5 demonstrates consistently strong performance across all modalities, with notable superiority in X-ray interpretation. GPT-5-mini maintains

competitive accuracy, particularly in MRI and CT. In contrast, GPT-5-nano and GPT-4o exhibit greater performance variability, with reduced accuracy in CT and MRI, suggesting their limited generalizability across complex modality-specific imaging patterns.

| Modality | GPT-5 (vs 4o) | GPT-5-mini | GPT-5-nano | GPT-4o |
|---|---|---|---|---|
| CT | 85.81% (↑9.33%) | 81.14% | 74.79% | 76.48% |
| MRI | 84.65% (↑10.97%) | 82.89% | 75.00% | 73.68% |
| X-ray | 94.74% (↑14.41%) | 86.98% | 80.33% | 80.33% |

**Figure 4.** Comparison of model accuracy across different imaging modalities, including CT, MRI, and X-ray. GPT-5 demonstrates consistently strong performance across all modalities, with notable superiority in X-ray interpretation. GPT-5-mini maintains competitive accuracy, particularly in MRI and CT. In contrast, GPT-5-nano and GPT-4o exhibit greater performance variability, with reduced accuracy in CT and MRI, suggesting their limited generalizability across complex modality-specific imaging patterns.

### 3.5 Evaluation on Radiology Visual Question Answering (VQA-RAD)

To rigorously evaluate the multimodal reasoning capabilities of GPT-5 and its variants in real-world clinical settings, we conducted experiments on the VQA-RAD benchmark—a challenging dataset specifically designed for radiology-oriented visual question answering (VQA). This task requires models to interpret medical images (such as X-rays and CT scans) in conjunction with domain-specific natural language queries, and to provide accurate, clinically meaningful answers. Unlike traditional image classification, success on VQA-RAD depends heavily on high-level reasoning, contextual understanding, and alignment between visual and textual modalities.

Under a zero-shot setting with no task-specific fine-tuning or prompt engineering, GPT-5 achieved a remarkable accuracy of 74.90%, decisively outperforming all baseline models. In contrast, GPT-4o reached only 69.91%, indicating a performance gap of +4.99%, despite its comparable scale and general-purpose capabilities. The performance drop is even more pronounced for lightweight variants: GPT-5-mini achieved 70.92%, and GPT-5-nano lagged behind with 65.34%. These results reveal that while smaller models may offer computational efficiency, they suffer from significant degradation in complex clinical reasoning tasks that require deeper cross-modal comprehension.

Importantly, GPT-5's superior performance underscores its enhanced alignment between radiological image features and medical linguistic priors. Even without task-specific adaptation, it demonstrates a robust understanding of both anatomy and clinical semantics, such as lesion interpretation, modality recognition, and procedural context.

These findings suggest that GPT-5 is not only more accurate but also more clinically reliable for high-stakes medical VQA applications. Its consistent edge in performance—particularly under zero-shot conditions—further reinforces its potential for downstream clinical decision support, automated radiology reporting, and interactive diagnostic assistants.

### 3.6 Evaluation on Medical Physics Board Exam

To further assess the models' performance on clinically relevant knowledge, we evaluated all four GPT variants using the in-house curated *Medical Physics Board Exam* dataset. This dataset comprises 150 high-quality multiple-choice questions modeled after standardized certification exams in radiation oncology physics. The questions span a broad range of topics, including treatment planning systems, dosimetry, beam modeling, image-guided radiation therapy, quality assurance, and radiation protection. Each item contains four or five answer options with a single correct choice, designed to test not only factual recall but also conceptual understanding, numerical reasoning, and applied physics knowledge.

Among the evaluated models, GPT-5 demonstrated the strongest performance, correctly answering 136 out of 150 questions, yielding an impressive accuracy of 90.7%. GPT-5-mini followed closely, achieving 86.7% accuracy with 130 correct answers. GPT-4o, while slightly less consistent, still achieved a respectable 83.3% (125 correct). GPT-5-nano, being the most lightweight variant, performed comparatively lower, correctly answering 110 questions, with an accuracy of 73.3%.

Notably, GPT-5's score surpasses the typical human passing threshold for the American Board of Radiology (ABR) Part I exam in medical physics, which is generally around 70% to 75%. This suggests that GPT-5 is capable of operating at or above the level expected of entry-level board-eligible clinical physicists. The fact that GPT-5 not only exceeds this benchmark but also outperforms smaller model variants by a considerable margin reinforces its utility in expert-level reasoning and decision-support tasks in radiation oncology.

## 4. Discussion

This study provides a first-of-its-kind evaluation of GPT-5 and its smaller variants in multimodal, domain-specific tasks relevant to radiology, radiation oncology, and medical physics. Across both visual question answering (VQA) datasets and the Medical Physics Board Examination–style assessment, GPT-5 consistently demonstrated higher accuracy than its predecessors and compact variants, indicating substantial improvements in multimodal reasoning, clinical concept grounding, and numerical problem-solving. Nevertheless, several findings warrant careful interpretation, and the results reveal both strengths and limitations that should inform future development and deployment.

GPT-5's consistent superiority over GPT-4o and smaller GPT-5 derivatives across SLAKE, VQA-RAD, and domain-specific physics questions suggests that architectural refinements and expanded training corpora have meaningfully improved domain reasoning capacity. Notably, the model's advantage in open-ended question formats points to improved generative precision, a critical requirement for free-text radiology reporting and complex treatment planning justifications. The strong performance on board-style physics questions—surpassing estimated human passing thresholds—further indicates that GPT-5 has acquired a level of procedural and conceptual knowledge that could support education, training, and decision augmentation in high-stakes clinical settings.

Despite its overall advantage, GPT-5's gains were not uniform. Performance gains were largest in anatomically complex regions (e.g., thoracic mediastinum, brain), but narrower in tasks involving more straightforward content categories (e.g., modality or plane recognition), where smaller models occasionally approached parity. Such heterogeneity suggests that GPT-5's improvements may stem from enhanced high-level reasoning rather than uniformly better perceptual feature alignment. This aligns with prior work showing that large multimodal LLMs excel most when tasks require integration of heterogeneous signals and reasoning chains, rather than purely visual classification.

The study's zero-shot evaluation protocol, while providing a clear measure of generalization, does not account for domain-specific adaptation. It remains possible that targeted fine-tuning, prompt engineering, or retrieval-augmented generation could close the gap for smaller models or further extend GPT-5's lead. Moreover, the datasets—while representative—cannot capture the full variability of real-world clinical images, language styles, and edge-case scenarios. For example, VQA-RAD and SLAKE, though well-established, contain relatively clean and curated examples that may underestimate the challenge posed by incomplete, noisy, or artifact-laden clinical data.

A critical barrier to deployment remains the absence of guarantees around factual correctness and reasoning transparency. Even when overall accuracy is high, isolated errors in high-stakes domains can have severe consequences. Our qualitative inspection revealed instances where GPT-5 generated plausible but factually incorrect responses, sometimes accompanied by overconfident explanations—a phenomenon consistent with "hallucination" patterns reported in other clinical LLM studies. The lack of explicit uncertainty estimates further complicates safe integration into workflows where human oversight is variable.

Several avenues emerge for extending this work. First, evaluations should incorporate more diverse and unstructured multimodal inputs, including dynamic imaging (e.g., cine MRI), time-series dosimetry data, and multilingual clinical narratives. Second, rigorous robustness testing under distribution shifts—such as scanner variability, protocol differences, or intentionally perturbed inputs—would clarify real-world resilience. Third, integrating calibrated uncertainty measures, explanation mechanisms, and user-in-the-loop reinforcement could address interpretability and trust. Finally, longitudinal studies examining how such systems affect clinical decision-making, training outcomes, and error rates in realistic settings will be essential to responsibly transition from benchmarking to deployment.

## 6. Conclusion.

Our findings indicate that GPT-5 represents a meaningful step forward in multimodal clinical AI capability, outperforming smaller variants and earlier-generation models across diverse medical imaging and physics tasks. However, realizing its potential in operational radiology and radiation oncology workflows will require targeted adaptation, robust safety mechanisms, and transparent evaluation under realistic conditions. The results here should be viewed as a baseline—one that underscores both the promise and the caution necessary when translating foundation model advances into safety-critical domains.